\documentclass[preprint]{elsarticle}
\usepackage[latin9]{inputenc}
\usepackage{amsmath}
\usepackage{amssymb}
\usepackage{graphicx}
\usepackage{esint}

\makeatletter

\journal{Journal of \LaTeX\ Templates}

\makeatother

\begin{document}
\begin{frontmatter}

\title{Finite-size effects on the minimal conductivity in graphene with
Rashba spin-orbit coupling}


\author[BME,MTA]{Péter Rakyta}


\author[mymainaddress]{László Oroszlány}


\author[mysecondaryaddress]{Andor Kormányos}


\author[mymainaddress]{József Cserti\corref{mycorrespondingauthor}}

\cortext[mycorrespondingauthor]{Corresponding author}

\ead{cserti@elte.hu}

\address[BME]{Department of Theoretical Physics, Budapest University of Technology
and Economics, H-1111 Budafoki út. 8, Hungary}

\address[MTA]{MTA-BME Condensed Matter Research Group, Budapest University of
Technology and Economics, H-1111 Budafoki út. 8, Hungary}

\address[mymainaddress]{Department of Physics of Complex Systems,E{ö}tv{ö}s University,H-1117
Budapest, Pázmány P{é}ter s{é}tány 1/A, Hungary}

\address[mysecondaryaddress]{Department of Physics, University of Konstanz, D-78464 Konstanz,
Germany}

\begin{abstract}
We study theoretically the minimal conductivity of monolayer graphene
in the presence of Rashba spin-orbit coupling. The Rashba spin-orbit
interaction causes the low-energy bands to undergo trigonal-warping
deformation and for energies smaller than the Lifshitz energy, the
Fermi circle breaks up into parts, forming four separate Dirac cones.
We calculate the minimal conductivity for an ideal strip of length
$L$ and width $W$ within the Landauer--Büttiker formalism in a continuum and in a tight binding model. 
We show that the minimal conductivity depends on the relative
orientation of the sample and the probing electrodes due to the interference of states related to 
different Dirac cones. 
We also explore the effects of finite system size
and find that the minimal conductivity can be lowered compared
to that of an infinitely wide sample. 
\end{abstract}
\begin{keyword}
\textbf{mesoscopic systems, quantum wires, carbon nanostructures, charge- and spin- transport } 
\end{keyword}
\end{frontmatter}


\section{Introduction}

More than half a century has passed since Landauer derived a formula
for the conductance of two terminal coherent devices~\cite{Landauer-1:cikk}.
Then 25 years ago \textit{Markus Büttiker} realized that the two terminal
Landauer formula can be extended to multi-terminal devices~\cite{PhysRevLett.65.2901}.
Now, in the literature this approach is commonly called Landauer--Büttiker
formalism. Over the years it become the standard tool for investigating
various quantum systems in nanophysics (for a review see Refs.~\cite{Carlo-Houten,Datta,Heinzel:book,Schon1}).
This approach has become an integral part of theoretical investigations
of modern solid states systems such as graphene~\cite{Novoselov_graphene-1}.
In the last decade different types of graphene nanostructures proved
to be one of the most technologically promising and theoretically
intriguing solid state systems. The dynamics of low energy excitations
in graphene is governed by an effective Hamiltonian corresponding
to massless two dimensional Dirac fermions. Hence many physical quantities 
such as the conductivity, the quantized Hall response and optical properties
are markedly different from those of conventional two dimensional
electron systems~\cite{neto:109}. In bilayer graphene, the interlayer hopping 
results in a trigonally warped Fermi surface which breaks 
up into four separate Dirac cone at low energies. The signatures of
this novel electronic structure has been studied first experimentally
by Novoselov \emph{et al.}~\cite{Novoselov-bilayer:cikk} and theoretically
by McCann and Fal'ko~\cite{mccann:086805}.

Graphene samples, despite the vanishing density of states, show a finite
conductivity at the charge neutrality point (at zero Fermi energy).
This feature of massless Dirac fermions, referred to as minimal conductivity,
was intensively studied with the Landauer--Büttiker formalism~\cite{PhysRevLett.96.246802,Katsnelson_Klein:ref,PhysRevB.85.041402}.
An alternative approach based on the Kubo formula has also been applied
to study this phenomenon in both monolayer and bilayer graphene~\cite{PhysRevLett.97.266802,PhysRevLett.99.066802}.
It was shown that in monolayer graphene for wide and short junction
the value of the minimal conductivity is $\sigma_{0}=\frac{4}{\pi}\frac{e^{2}}{h}$~\cite{PhysRevLett.96.246802,PhysRevB.85.041402}.
For bilayer graphene neglecting trigonal warping the conductivity
is $\sigma=2\sigma_{0}$, while including splitting of the Dirac cone
due to trigonal warping gives extra contributions to the conductivity,
increasing it to $\sigma=6\sigma_{0}$~\cite{PhysRevLett.99.066802}.
Later, for finite size of bilayer graphene it was shown by Moghaddam
and Zareyan~\cite{PhysRevB.79.073401} that the trigonal warping
results in an anisotropic behavior of the minimal conductivity.

Rashba spin-orbit (RSO) interaction arises once the mirror symmetry
of the bulk graphene sample is broken by the substrate or an applied
electric field perpendicular to the graphene sheet. The strength $\lambda$
of the RSO coupling is proportional to this electric field. Photoemission
experiments on graphene/Au/Ni(111) heterostructure revealed $\lambda\sim4$
meV~\cite{PhysRevLett.101.157601}. Recently, a strong Rashba 
effect with spin-orbit splitting of 70 meV has also been observed 
for graphene on Fe(110)~\cite{Varykhalov_nature_comm:cikk}. 

Enhanced RSO interaction has a major impact on the transport properties
of graphene derived samples. Recently the transfer matrix method has
been employed to study spin dependent transport properties of monolayer
graphene in the presence of inhomogeneous RSO coupling~\cite{Hasanirokh2014227,Razzaghi201589}.
An important consequence of the RSO interaction is that the low-energy
behavior of electrons in monolayer graphene with RSO coupling 
is related to that of bilayer graphene with trigonal warping but
without RSO interaction~\cite{PhysRevB.82.113405}. Therefore, we
expect that the minimal conductivity of monolayer graphene with RSO
interaction shows a similar anisotropic behavior as that obtained
for bilayer graphene in Ref.~\cite{PhysRevB.79.073401}. 

\begin{figure}
\centering \includegraphics[width=0.65\textwidth]{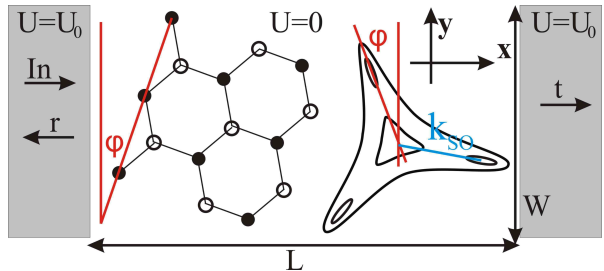}
\caption{Geometry of a graphene device of length $L$ and width $W$ between
two electrodes doped by potential $U_{0}$. Electrons incoming from
the left lead are reflected with amplitudes $r$ and transmitted with
amplitudes $t$. Between the two contacts we depict the real space
structure of the monolayer graphene flake (left side) and the energy
contours in reciprocal space around the $\mathbf{K}$ point. The zig-zag
direction of the graphene flake makes an angle $\varphi$ with the
electrode interface ($y$ direction).}
\label{fig_geometry:fig} 
\end{figure}

To see this anisotropic behavior, we calculate the minimal conductivity
using tight binding (TB) calculations and compare it to results obtained
from a continuous model. We study the effects of finite sample sizes
and the crystallographic orientation as well as the length dependent
oscillatory behavior of the minimal conductivity. In our two-terminal
calculations, the ballistic scattering region of monolayer graphene
with length $L$ and width $W$ is contacted by two highly doped regions
oriented at angle $\varphi$ with respect to the zig-zag direction
of the graphene lattice (see Fig.~\ref{fig_geometry:fig}). 
Doping in the electrodes is achieved by shifting the Fermi energy
with a large potential $U_{0}$ as it is commonly done in the literature
(see, \textit{e.g.}, Ref.~\cite{PhysRevLett.96.246802}).

\section{Landauer--Büttiker formalism for calculating the conductivity }

In the Landauer-Büttiker approach the conductance of a sample is given
by the transmission probabilities of an electron passing through it:
\begin{equation}
G=\frac{e^{2}}{h}\sum\limits _{m,n}|t_{mn}|^{2},\label{eq:buttiker_sigma}
\end{equation}
where $t_{mn}$ are the transmission amplitudes between the propagating
modes $n$ and $m$ of the left and right electrodes. 
In what follows, we calculate
the minimal conductivity in the TB model (for finite $W$)
and compare the results to that obtained in the continuous model (for
$W\to\infty$). Both in TB and continuous model the transmission
amplitudes $t_{mn}$ are calculated by solving the scattering problem
of the system. 
Then the minimal conductivity is defined as $\sigma=\frac{L}{W}\,G$,
with the conductance $G$ calculated from Eq.~(\ref{eq:buttiker_sigma})
at the charge neutral point of graphene, ie, at $E_{F}=0$.

\subsection{Tight binding model of graphene including RSO coupling}

\label{TB:sec}

\begin{figure}
\centering \includegraphics[width=6cm]{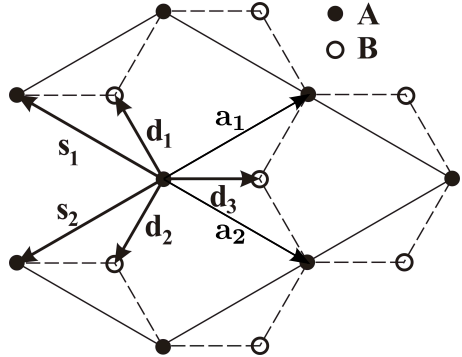} 
\caption{Geometry of a graphene sheet. The unit vectors of the hexagonal lattice
are $\textbf{a}_{1}$ and $\textbf{a}_{2}$, while $\textbf{d}_{1}=(\textbf{a}_{2}-2\textbf{a}_{1})/3$,
$\textbf{d}_{2}=(\textbf{a}_{1}-2\textbf{a}_{2})/3$ and $\textbf{d}_{3}=(\textbf{a}_{1}+\textbf{a}_{2})/3$
are vectors pointing to the neighboring atoms.}
\label{fig:geometry} 
\end{figure}

In the TB model the Hamiltonian $H_{TB}$ of monolayer
graphene with RSO coupling can be written as~\cite{PhysRevLett.95.226801,PhysRevB.82.113405}
\begin{subequations} \label{Ham_TB:eq} 
\begin{align}
H_{TB} & =H_{0}+H_{R},\hspace{7mm}\text{where}\\
H_{0} & =-\gamma\sum\limits _{\left\langle i,j\right\rangle ,\sigma}\left(a_{i\sigma}^{\dagger}b_{j\sigma}+{\rm h.c.}\right),\\
H_{R} & =i\,\lambda\sum\limits _{\left\langle i,j\right\rangle ,\mu,\nu}\left[a_{i\mu}^{\dagger}\left(\boldsymbol{s}_{\mu\nu}\times\mathbf{\widehat{d}}_{\left\langle i,j\right\rangle }\right)_{z}b_{j\nu}-h.c.\right].
\end{align}
\end{subequations}Here $H_{0}$ is the Hamiltonian of bulk graphene
sheet taking into account only nearest neighbor hopping, with hopping
amplitude $\gamma$. The operator $a_{i\sigma}^{\dagger}$ ($a_{i\sigma}$)
creates (annihilates) an electron in the $i$th unit cell with spin
$\sigma$ on sublattice $A$, while $b_{j\sigma}^{\dagger}$ ($b_{j\sigma}$)
has the same effect on sublattice $B$ and $\textrm{h.c.}$ stands
for hermitian conjugate. The unit cell is given by the unit vectors
$\mathbf{a}_{1}$ and $\mathbf{a}_{2}$ as shown in Fig.~\ref{fig:geometry}.
The Hamiltonian $H_{R}$ describes the Rashba spin-orbit interaction
where $\boldsymbol{s}=(s_{x},s_{y},s_{z})$ are the Pauli matrices
representing the electron spin, and $\mu,\nu=1,2$ denote the $\mu\nu$
matrix elements of the Pauli matrices. Here vectors $\mathbf{d}_{\left\langle i,j\right\rangle }$
connect the nearest neighbor atoms $\left\langle i,j\right\rangle $
pointing from $j$ to $i$ as shown in Fig.~\ref{fig:geometry},
and $d$ is the distance between them, and 
$\mathbf{\widehat{d}}_{\left\langle i,j\right\rangle }=\mathbf{d}_{\left\langle i,j\right\rangle }/d$
are unit vectors. 

The strength of the spin-orbit coupling is denoted by $\lambda$ which
may arise due to a perpendicular electric field or interaction with
a substrate.

Using the standard Green's function techniques~\cite{TB_Green,PhysRevB.78.035407,PhysRevB.90.125428}
based on the Landauer--Büttiker approach we calculate the transmission
amplitudes for armchair and zig-zag orientation of the sample.

\subsection{Continuous model of graphene including RSO coupling}

\label{cont:sec}

The Hamiltonian of the continuous model as a long wave approximation
of the TB Hamiltonian $H_{TB}$ in Eq.~(\ref{Ham_TB:eq})
describes low energy excitations around the $\mathbf{K}$ and $\mathbf{K}^{\prime}$
points. In our previous publication~\cite{PhysRevB.82.113405} we
showed that starting from the tight-binding Hamiltonian suggested
in Ref.~\cite{PhysRevLett.95.226801} to describe RSO coupling in monolayer
graphene one can arrive at a form of the Hamiltonian that is unitary equivalent
to that of bilayer graphene without RSO interaction but including
the trigonal warping effect due to interlayer hopping~\cite{mccann:086805,PhysRevLett.99.066802}.
In the continuous model the Hamiltonian $H_{K}$ at the $\mathbf{K}$
point of the Brillouin zone (BZ) reads as: 
\begin{equation}
H_{K}=\begin{pmatrix}0 & v_{F}p_{-} & 0 & v_{\lambda}p_{+}\\
v_{F}p_{+} & 0 & -3i\lambda & 0\\
0 & 3i\lambda & 0 & v_{F}p_{-}\\
v_{\lambda}p_{-} & 0 & v_{F}p_{+} & 0
\end{pmatrix}
\label{eq:HB}
\end{equation}
where $v_{F}=3\gamma d/(2\hbar)$, $v_{\lambda}=3\lambda d/(2\hbar)$,
$p_{\pm}=p_{x}\pm ip_{y}$ and $p_{x},p_{y}$ are momentum operators.
The Hamiltonian $H_{K}$ in Eq.~(\ref{eq:HB}) is written in the
basis ${(|A\uparrow\rangle,|B\uparrow\rangle,|A\downarrow\rangle,|B\downarrow\rangle)}^{T}$
where $\{\uparrow,\downarrow\}$ refer to spin orientations. 
A unitary equivalent result can be obtained around the Dirac point $\mathbf{K}^{\prime}$.
The four eigenvalues of the Hamiltonian (\ref{eq:HB}) as a function of the
wave number $\mathbf{k} = (k_x,k_y) = k(\cos\alpha,\sin\alpha)^T$ are given by
\begin{subequations}
\begin{align}
 E_n^{\pm}(\mathbf{k}) &= \pm \hbar v_F\,
\sqrt{\frac{1}{2}\left[k_\lambda^2 + k^2 \left(2+\beta^2\right) + (-1)^n \sqrt{\Upsilon}\right] }, \,\,\, \textrm{with} \\
\Upsilon &= k_\lambda^4 +  2k^2 k_\lambda^2 (2-\beta^2) + k^4 \beta^2 (4+\beta^2) -
 8 k^3 k_\lambda \beta \sin (3 \alpha),
\end{align}
where $\beta=v_{\lambda}/v_F = \lambda/\gamma$ is the dimensionless strength of the spin-orbit coupling,
$k_\lambda =  2 \beta /d$ and $n=1,2$.
\end{subequations}%

\begin{figure}[hbt]
\centering \includegraphics[width=12cm]{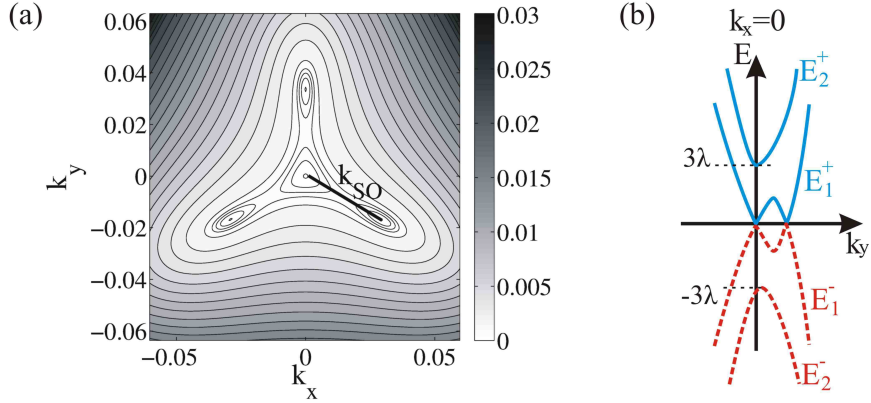} 
\caption{(a) Contour plot of the positive and low-energy band $E_{1}^{+}$ (in units of $\lambda$)
around the $\mathbf{K}$ point for $\beta=0.034$. Wave vector components
$k_{x}$, $k_{y}$ are in units of $3\lambda/(\hbar v_{F})$. The
center of the pockets has a $2\pi/3$ rotational symmetry. The distance
between the center of the pockets and the central Dirac points (points
$\mathbf{K}$) is $k_{\textrm{SO}}$ given in the text. 
(b) The four energy bands along the direction $k_y$ with $k_x=0$. 
\label{fig:spectrum}}
\end{figure}

Figure~\ref{fig:spectrum} shows the contour plot of the positive and low-energy band $E_{1}^{+}$ and the spectrum along the $k_y$ direction. 
The spectrum has a threefold symmetry similar to that of bilayer graphene.
At moderate energy, direct hopping between $\Psi_{A\uparrow}$ and
$\Psi_{B\downarrow}$ leads to trigonal warping of the constant energy
lines about each valley, but at an energy $E$ less than the Lifshitz
energy $E_{L}=\gamma\beta^{2}/(4+\beta^{2})$ the effect
of trigonal warping is dramatic. It leads to a Lifshitz transition~\cite{Abrikosov:book}:
the constant energy line is broken into four pockets, which we refer
to as central and three leg parts. The Fermi surface is approximately
triangle like in the central part and each leg part it is elliptical. The
distances of the center of the leg parts from the $\mathbf{K}$ point
are $k_{\textrm{SO}}=2\beta^{2}/d$ (see Fig.~\ref{fig:spectrum}).

As it has been shown in our previous work~\cite{PhysRevLett.99.066802}
the Lifshitz transition strongly affects the transport properties
of monolayer graphene as well. The anisotropy of the minimal conductivity
in bilayer graphene related to the interference effects between the
leg parts was predicted by Moghaddam \emph{et al.}~\cite{PhysRevB.79.073401}.
Therefore, in monolayer graphene including the RSO interaction, we
also expect a strong anisotropy in its conductivity depending on the
orientation of the leg parts with respect to the electrodes. To see
this we calculate the transmission probabilities $t_{mn}$ in Eq.~(\ref{eq:buttiker_sigma})
and the minimal conductivity by solving the scattering problem. If
we consider the short and wide junction limit ($W\gg L$), then the
electronic states can be specified by their energy $\varepsilon$
and the transverse wavenumber $q$ which are conserved during the
scattering process. For a given $\varepsilon$ and $q$ there are
four solutions for the longitudinal wave vector $k^{l}$ which satisfies
the characteristic equation $\det\left[H_{K}(k^{(l)},q)-I_{4}\,\varepsilon\right]=0$,
where $I_{4}$ is the $4\times4$ identity matrix. Electronic states
in the scattering region ($0\leq x\leq L$) are denoted by 
$\Psi_{\text{sc}}^{l}(q)=\Phi_{\text{sc}}^{l}e^{i(k^{l}x+qy)}$,
where $\Phi_{\text{sc}}^{l}$ satisfy relation for all possible quantum
numbers $l$: 
\begin{equation}
H_{K}(k^{l},q)\,\Phi_{\text{sc}}^{l}=\varepsilon\,\Phi_{\text{sc}}^{l}.\label{eq:eigvec}
\end{equation}
The scattering state between the electrodes is then a linear combination
of these four electronic states. The longitudinal wave numbers $k_{L/R}^{n}$
and the corresponding electronic states $\Psi_{L/R}^{n}(q)=\Phi_{L/R}^{n}\,e^{i(k_{L/R}^{n}x+qy)}$
in the left (L) and right (R) leads can be obtained analogously with
a substitution $\varepsilon\rightarrow\varepsilon-U_{0}$ (here $U_{0}$
is the potential on the left and right electrodes as indicated in
Fig.~\ref{fig_geometry:fig}. If we assume an incident states in
the L electrode, than the resulted scattering state can be written
in the form: 
\begin{equation}
\Psi^{n}(q)=\left\{ \begin{array}{ll}
\Psi_{L}^{n,\rightarrow}(q)+\sum\limits _{n'=1}^{2}r_{n'n}(q)\Psi_{L}^{n',\leftarrow}(q), & x<0,\\
\sum\limits _{l=1}^{4}A_{p}\Psi_{\text{sc}}^{(l)}(q), & 0\leq x\leq L,\\
\sum\limits _{m=1}^{2}t_{mn}(q)\Psi_{R}^{m,\rightarrow}(q), & L<x,
\end{array}\right.
\end{equation}
where $r_{n'n}(q)$ and $t_{mn}(q)$ are the reflection and transmission
amplitudes, and we introduced the arrow $\rightarrow$ ($\leftarrow$)
to label the right (left) propagating electron states in the leads.
The reflection and transmission amplitudes have to be determined (together
with coefficients $A_{n}$) by imposing the continuity condition of
the wave functions at the interfaces $x=0$ and $x=L$.

Finally, inserting the transmission probabilities $t_{mn}(q)$ into
Eq.~(\ref{eq:buttiker_sigma}) we find the conductance $G$. The
summation over the transverse wave numbers is replaced in a good approximation
by the integration $\frac{W}{2\pi}\int\textrm{d}q$. Then the minimal
conductivity reads: 
\begin{equation}
\sigma=2\,\frac{L}{W}\,G=\frac{\sigma_{0}}{4}\,L\int\limits _{-\infty}^{\infty}{\rm d}q\sum\limits _{m,n}\left|t_{mn}(q)\right|^{2},\label{eq:G}
\end{equation}
where in the first equation the factor 2 corresponds to the valley
degeneracy.

\section{Results: the minimal conductivity of monolayer graphene with RSO
interaction}

\begin{figure}
\centering \includegraphics[width=9.5cm]{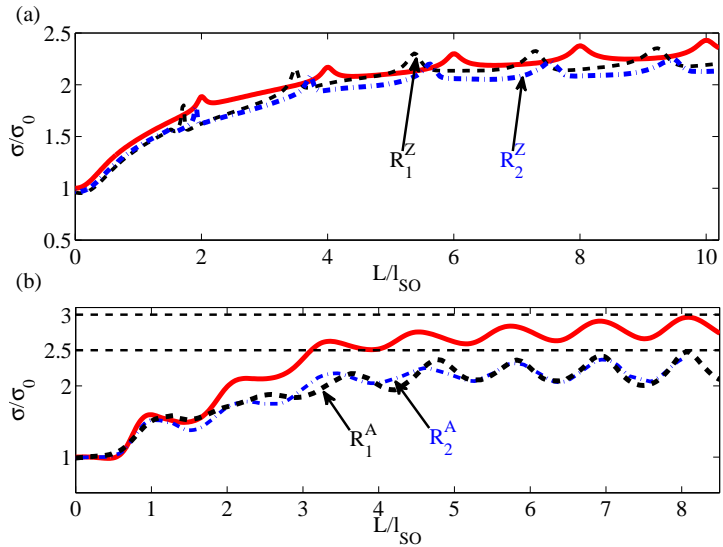}
\caption{ (Color online) The conductivity (in units of $\sigma_{0}$) of the
junction as a function of length $L$ (in units of $l_{SO}$) for
(a) zigzag and (b) armchair orientation in continuous model obtained
from Eq.~(\ref{eq:G}) (red solid lines) and from TB calculation
(blue dash-dotted and black dashed lines) with two different aspect ratios $R_{1}^{Z}=W/L=4.71$
and $R_{2}^{Z}=W/L=6.74$ for zigzag orientation, and $R_{1}^{A}=W/L=3.12$
and $R_{2}^{A}=W/L=5.80$ for armchair orientation. 
The two horizontal dashed lines in Fig.~b represent 
the upper limit of the conductivity calculated 
from the continuous ($3 \sigma_0$) and TB model ($\frac{5}{2}\, \sigma_0$)
as described in the text.
\label{fig:conductance}}
\end{figure}

The minimal conductivity as function of $L$ obtained from the continuous
model and from TB calculations for zig-zag and armchair
orientation are shown in Fig.~\ref{fig:conductance}. 

As described in Ref.~\cite{PhysRevB.79.073401} the RSO interaction
can be characterized by a length scale $l_{SO}=\pi/k_{\textrm{SO}}\sim1/\lambda^{2}$.
For short junctions or at low RSO coupling $\lambda$, that is in
the limit $L/l_{SO}\to0$, the conductivity for both the armchair
and zig-zag orientation starts with $\sigma(L/l_{SO}=0)=\sigma_{0}$.
Increasing $L/l_{SO}$ the conductivity calculated from the continuous
model tends to $\sigma=3\sigma_{0}$ and $\sigma=7/3\,\sigma_{0}$
for the armchair and zig-zag orientation, respectively. 

In the TB calculation for zig-zag orientation, depicted
in Fig.~\ref{fig:conductance}a, the conductivity closely follows
that of the continuous model and tends towards $\sigma_{TB}=7/3\,\sigma_{0}$
for longer junctions. Increasing the $W/L$ ratio the subtle peaks
of the TB and continuous models approach each other. On
the other hand, for the armchair orientation, shown in Fig.~\ref{fig:conductance}b, 
the results of the TB calculation and the continuous
model start to deviate for $L/l_{SO}\gtrapprox1.1$, that is for
increased RSO coupling $\lambda$, tending to a markedly lower value
$\sigma_{TB}=5/2\,\sigma_{0}$. We also observe an enhanced oscillatory
behavior as the function of $L/l_{SO}$ as compared to the calculation
done in the zig-zag direction. 

\begin{figure}[hbt]
\centering \includegraphics[width=9cm]{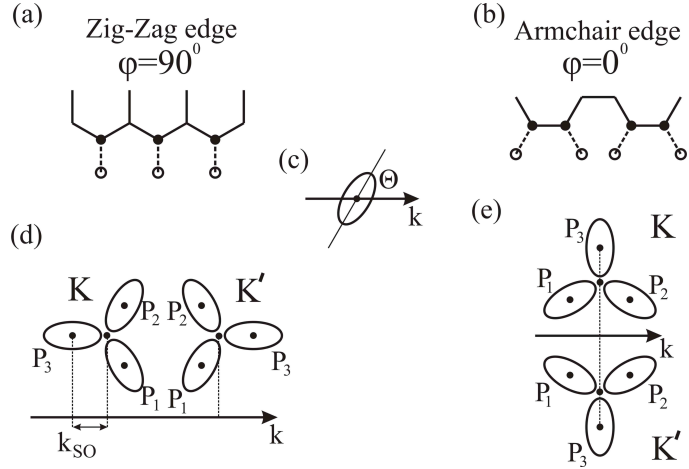} 
\caption{Schematic drawing of (a) zig-zag and (b) armchair edges. Empty circles
and dashed lines correspond to the nearest missing sites and bonds
to the edge of the ribbon. (c) The orientation of one pocket (an ellipse)
is given by the angle $\Theta$ between the propagation direction
($k$ axis) and the semi-major axis of the ellipse. (d) Orientation
of the pockets $P_{1}$, $P_{2}$, and $P_{3}$ for zig-zag and (e)
for armchair edges in the Brillouin zone.}
\label{fig:edges} 
\end{figure}

To understand this behavior of the conductivity it is instructive
to consider the orientation of the Fermi surface around the $\mathbf{K}$
and $\mathbf{K}^{\prime}$points with respect to the direction of
propagation as shown in Fig.~\ref{fig:edges}. As noted before, trigonal
warping due to the RSO interaction brakes the Fermi surface into a
central pocket (dots in Figs.~\ref{fig:edges}d and~\ref{fig:edges}e at the $\mathbf{K}$
and $\mathbf{K}^{\prime}$points) and three extra leg pockets labeled
by $P_{1}$, $P_{2}$ and $P_{3}$. 

First we explain the oscillatory behavior of the conductivity shown
in Fig.~\ref{fig:conductance}b for armchair orientation. In Fig.~\ref{fig:edges}e
the zero energy modes both around the $\mathbf{K}$ and $\mathbf{K}^{\prime}$
point are at the center of pocket $P_{1}$, $P_{2}$ and $P_{3}$,
and at the center of the isotropic Dirac cone. Out of these four modes
two (the central\textbf{ }Dirac cone and pocket $P_{3}$) have a wave
number $k=0$ (along the propagating direction) and for the other
two modes the wave numbers are $k=\pm\frac{\sqrt{3}}{2}k_{\textrm{SO}}$
(the centers of pocket $P_{1}$ and $P_{2}$ in Fig.~\ref{fig:edges}e.
The latter two non-zero propagating modes explain the oscillatory
behavior of the conductivity shown in Fig.~\ref{fig:conductance}b.
The phase shift between the finite $k$ propagating modes (accumulated
over one period) for an electron bouncing between the electrodes is
$\Delta\Phi_{m}=\pm m\sqrt{3}k_{\textrm{SO}}L$, where $m=0,1,2$.
Then the shortest period of the conductivity is given by $\Delta\Phi_{1}=2\pi$
from which one finds 
\begin{equation}
\frac{L}{l_{SO}}=\frac{2}{\sqrt{3}}\approx1.15.
\end{equation}
This periodicity can be clearly seen in Fig.~\ref{fig:conductance}b 
for both TB and continuous cases.

In the case of the zig-zag orientation all pockets are centered at
finite $k$. In both valleys the centers of the central pocket and
pocket $P_{3}$ are separated by $k_{SO}$. this gives $L/l_{SO}=2$
as the shortest modulation period in agreement with our data presented
in Fig.~\ref{fig:conductance}a.

Now we explain the marked discrepancy between the continuous model
and the TB calculations performed in the armchair orientation.
For strong RSO interaction the conductivity calculated in the tight
binding approach can be estimated as follows. In general each pocket
$P_{1},P_{2}$ and $P_{3}$ shown in Fig.~\ref{fig:edges} corresponds
to one anisotropic Dirac cone and gives a separate contribution to
the conductivity. The total conductivity is given by 
\begin{subequations}
\begin{align}
\sigma & =n_{C}\sigma_{C}+\sum_{i}n_{i}\sigma(\Theta_{i}),\,\,\,\text{where}\\
\sigma(\Theta_{i}) & =\frac{v_{a}^{2}\cos^{2}(\Theta_{i})+v_{b}^{2}\sin^{2}(\Theta_{i})}{v_{a}v_{b}}\frac{\sigma_{0}}{4},
\end{align}%
\label{total_sigma:eq}%
\end{subequations}%
and $\sigma_{C}=\sigma_{0}/4$
is the contribution from the central Dirac cone~\cite{PhysRevLett.96.246802}
while $\sigma(\Theta_{i})$ is the minimal conductivity related to a single anisotropic Dirac cone. 
This result was first derived
by Nilsson et al.~in Ref.~\cite{PhysRevB.78.045405}. Note that
the same result can be obtained by the general approach developed
in Ref.~\cite{PhysRevB.85.041402}. Here $n_{C}$ and $n_{i}$ are
the number of open channels for the central Dirac cone and the leg pocket $P_{i}$, 
respectively, and $v_{a}$ and $v_{b}$ are the
Fermi velocities along the two principal axes of the ellipse corresponding
to the pocket $P_{i}$ with $i=1,2,3$. For our case $v_{b}=3v_{a}$~\cite{PhysRevB.78.045405}
for the three legs. 
$\Theta_{i}$ is the angle of the direction of the semi-major axis
of the ellipse with respect to the direction of propagation (see Fig.~\ref{fig:edges}c).
One can see from Fig.~\ref{fig:edges} that around the $\mathbf{K}$
point for armchair orientation $\Theta_{i}=7\pi/6;11\pi/6;\pi/2$
for pocket $P_{1}$, $P_{2}$ and $P_{3}$, respectively, while for
zig-zag orientation $\Theta_{i}=5\pi/3;\pi/3;\pi$ around the $\mathbf{K}$
point, and $\Theta_{i}=4\pi/3;2\pi/3;0$ around the $\mathbf{K}^{\prime}$
point for the pocket $P_{1}$, $P_{2}$ and $P_{3}$, respectively
(see Fig.~\ref{fig:edges}).

We now determine the number of open channels $n_{C}$ and $n_{i}$
in Eq.~(\ref{total_sigma:eq}).\textbf{ }The boundary condition for
the system demands that the wave function at the two edges of the
ribbons should be zero at the empty sites shown schematically in Figs.~\ref{fig:edges}a
and~\ref{fig:edges}b. Thus, including the spin we have \textit{four}
equations to satisfy the boundary conditions. For a given energy $\varepsilon$
and wave number $k$ corresponding to the propagating mode the possible
transverse modes can be calculated from the dispersion relation. Graphically
it means that these transverse modes can be obtained by drawing a
vertical line at a given $k$ that intersects the given constant energy
$\varepsilon$ contour. For example, for zig-zag orientation for wave
number $k$ for which the vertical line passes through the center
of pocket $P_{3}$ around the $\mathbf{K}$ point there are two transverse
modes (the vertical line crosses the energy contour at two points
in Fig.~\ref{fig:edges}d, while for wave number $k$ for which
the vertical line passes through the center of pockets $P_{1}$ and
$P_{2}$ we have four transverse modes. Hence, it follows that in
the first case the number of open channels $n_{1}=0$ since the four
boundary conditions cannot be satisfied by two transverse modes. Similarly,
for the central isotropic Dirac cone $n_{c}=0$. However, for the
second case $n_{1}=n_{2}=1$ because we have four transverse modes.
The same is true for the propagating mode $k$ around the $\mathbf{K}^{\prime}$
point (valley degeneracy). In summary, the open channels for zig-zag
ribbons are $n_{C}=0$ and $n_{1}=n_{2}=2,n_{3}=0$. From a similar
consideration we find that for armchair orientation $n_{C}=2$ and
$n_{1}=n_{2}=1,n_{3}=2$. Thus the minimal conductivity of monolayer
graphene with RSO coupling given by (\ref{total_sigma:eq}) is $\sigma=7/3\,\sigma_{0}$
for zig-zag and $\sigma=5/2\,\sigma_{0}$ for the armchair orientation,
in very good agreement with the TB calculations.

\section{Conclusions}

We have investigated the minimal conductivity of monolayer graphene
in the presence of Rashba spin-orbit interaction. We have employed
tight binding calculations and a continuous model, to determine the
interplay of the crystallographic orientation of the sample with the
anisotropic nature of the minimal conductivity. Contrasting the results
obtained for a graphene strip of finite width to that of an infinitely
wide sample, we show that the boundary condition for a finite flake
may, depending on the orientation, reduce the value of the minimal
conductivity compared to that of the infinitely wide. All our calculations
have been performed in the spirit of the Landauer-Büttiker approach.
We hope our results are a tribute for the long lasting legacy of 
this simple yet powerful formalism and to the memory of Markus Büttiker.

\section*{Acknowledgements}

The authors would like to thank A. Pályi for stimulating discussions.
This work was 
supported by the Hungarian Science Foundation OTKA under the contract
No.~108676.

\section*{References}


\end{document}